\documentclass[pdftex,iop]{emulateapj}
\journalinfo{\textsc{The Astrophysical Journal}}
\slugcomment{Received 10 January 2014, Accepted 18 June 2014}
\usepackage{url}
\usepackage{natbib}
\usepackage{booktabs}
\usepackage{epstopdf}

\usepackage[colorlinks,
        hypertexnames=false,
        pagebackref=true,
        linkcolor=blue,     
    citecolor=blue,        
    filecolor=magenta,      
    urlcolor=cyan           
]{hyperref}                 
\usepackage{lineno}
\usepackage{amsmath}
\newcommand{\Ha}{H$\alpha$}

\shorttitle{3D Magnetic Field and Flare Evolution}
\shortauthors{Vemareddy and Wiegelmann}

\begin{document}
\title{Quasi-Static 3D-Magnetic Field Evolution in Solar Active Region NOAA 11166 Associated with X1.5 Flare}
\author{P.~ Vemareddy$^1$, and T.~ Wiegelmann$^2$}
\affil{$^1$Udaipur Solar Observatory, Physical Research Laboratory,
Udaipur-313 001, India} \affil{$^2$Max-Planck-Institut f$\ddot{u}$r Sonnensystemforschung, G$\ddot{o}$ttingen-37077, Germany}
\email{vema@prl.res.in, wiegelmann@mps.mpg.de}


\begin{abstract}
We study the quasi-static evolution of coronal magnetic fields constructed from the Non Linear Force Free Field (NLFFF) approximation aiming to understand the relation between the magnetic field topology and ribbon emission during an X1.5 flare in active region (AR) NOAA 11166. The flare with a quasi-elliptical, and two remote ribbons occurred on March 9, 2011 at 23:13UT over a positive flux region surrounded by negative flux at the center of the bipolar AR. Our analysis of the coronal magnetic structure with potential and NLFFF solutions unveiled the existence of a single magnetic null point associated with a fan-spine topology and is co-spatial with the hard X-ray source. The footpoints of the fan separatrix surface agree with the inner edge of the quasi-elliptical ribbon and the outer spine is linked to one of the remote ribbons. During the evolution, the slow footpoint motions stressed the fieldlines along the polarity inversion line and caused electric current layers in the corona around the fan separatrix surface. These current layers trigger magnetic reconnection as a consequence of dissipating currents, which are visible as cusped shape structures at lower heights. The reconnection process reorganised the magnetic field topology whose signatures are observed at the separatrices/QSL structure both in the photosphere and corona during the pre-to-post flare evolution. In agreement with previous numerical studies, our results suggest that the line-tied footpoint motions perturb the fan-spine system and cause null point reconnection, which eventually causes the flare emission at the footpoints of the fieldlines.
\end{abstract}

\keywords{Sun:  reconnection--- Sun: flares --- Sun: magnetic fields---
Sun: topology --- Sun: fieldline ---Sun: evolution}




\section{Introduction}

\label{Intro}

Solar flares are caused by energy release in the corona due to magnetic reconnection. Flares emit electromagnetic radiation in a wide 
range of wavelengths and are frequently associated with coronal mass ejections and they expel solar energetic particles. Consequently they influence and drive space whether events (e.g., see \citealt{schrijver2011,vemareddy2011}). Bright ribbons are commonly observed in \Ha~and in EUV during both eruptive and non-eruptive flares. According to the standard 2D flare models (CSHKP; \citealt{carmichael1964, sturrock1966, hirayama1974, kopp1976}), these phenomena are caused by magnetic reconnection in current sheets. During these processes the charged coronal particles become energized and accelerated. Some particles gyrate around the fieldlines and propagate toward their foot points, precipitating at different layers of the solar atmosphere. They appear as ribbons in a wide range of wavelengths. Therefore, ribbons are nothing but the footpoints of coronal loops providing information about the linkage and fieldline topology in flaring active regions (AR). Such a proposed 2D picture of the fieldline connectivity is difficult to imagine and leads often to elusive interpretations.

Reconnection corresponds to the change of the magnetic connectivity by breaking and reconnecting fieldlines in resistive regions, where ideal magneto-hydrodynamics (MHD) is violated and the plasma is not frozen in the magnetic field. By tracing the temporal evolution of three-dimensional (3D) magnetic structures, it is possible to get insights into the conditions leading to an eruption. Previous topological analysis of magnetic structures constructed from analytic configurations or potential and linear force-free extrapolations revealed preferred sites for current sheet formation and associated features like null points, separatrix surfaces, and separator lines \citep{priest1989, demoulin1993, demoulin1997}. Magnetic null points \citep{lau1990,parnell1996} generate separatrices of two flux systems of individual connectivity. Across the separatrix surfaces the connectivity is discontinuous and contains an infinite magnetic field gradient. Since not all reconnecting configurations contain null points \citep{demoulin1997}, the concept of separatrices was generalized to quasi-separatrix layers (QSLs; \citealt{priest1992, priest1995}). Unlike separatrices, QSLs are locations with continuous connectivity, but still with large connectivity gradients. These are also favorable sites for the formation and build-up of current layers, similar to separatrices in the presence of footpoint motions.

Subsequently, it was important to search for the locations of such topological features in observational data and several reports gave evidences for the presence of magnetic null points as well as QSLs in flaring ARs \citep{mandrini1997,aulanier2000,fletcher2001, manoharan2003, xudong2012b, savcheva2012}. The role of these topological features for reconnection are explored in several theoretical \citep{demoulin1993,demoulin1996,demoulin1997, priest2000, longcope2005} and numerical \citep{moreno2008,pariat2009,masson2009} studies.

During the evolution of an AR, magnetic energy is built up (during several hours or even days) in the corona due to flux emergence and displacements \citep{schrijver2009}. For these slowly evolving processes, the characteristic travel time of Alfven waves through the whole region is orders of magnitude lower than the global evolution time of the region. Consequently, dynamical and inertial effects are negligible. Moreover, the plasma pressure and gravity are negligible as well, because they are small compared to magnetic pressure and tension, which counterbalance each other and lead to a vanishing Lorentz force. Therefore, the slow evolution of ARs can be approximated by a series of quasi-stationary, force-free equilibria in the low-plasma $\beta$ (ratio of plasma and magnetic pressures) corona. Modeling the magnetic configuration at each stage of the AR evolution in this scenario allows a detailed study of the connection between ribbons, topology, and energetics.

Nevertheless, the coronal field is force-free, the photospheric field is far away (e.g., \citealt{wiegelmann2006}) from that state. Furthermore, measurement errors make it difficult to reconstruct the coronal fields. Therefore, the force-free model is a sophisticated approximation of the coronal field within the limits of current observational capabilities. Continuous photospheric field measurements of the \textit{Solar Dynamic Observatory} with a cadence of 12 minutes are used as boundary condition to model the 3D coronal magnetic field. The resulting data cubes are analyzed  to reveal their topological features, related current distributions and to locate sites for energy release during explosive events like flares and coronal mass ejections. This study tries to address the possible role of magnetic topology for solar energetic events and the related flare emission.

In the present work, we analyze an X-class flare with a main quasi-elliptical ribbon and additional two elongated remote ribbons. Such elliptical geometries occur only for specific magnetic topologies and have been therefore studied rarely. To our knowledge, \citet{masson2009} reported for the first time a c-class flare with a circular ribbon accompanied by two extended remote ribbons in an emerging AR. The authors used MHD simulations in a line-tied boundary approximation to explain the formation of current sheets, the nature of reconnection and several other observational features like ribbon brightening, H$\alpha$ and EUV-emission. We use a similar approach to interpret the observations in this paper, but our analysis is based on a quasi-static force-free model instead of dynamic MHD simulations. 

We organize the paper as follows. In Section~\ref{Data} we describe the data and in Section~\ref{Methods} our coronal magnetic field model. We analyse the morphological flare evolution and the associated magnetic field properties in Section~\ref{FlareMorphEvol}, which is followed by an analysis of the magnetic structure, topology and coronal current distribution in Sections~\ref{MagStrTopo} and~\ref{RibCurrRec}. We emphasis the relevance of our results in comparison with recent theoretical and numerical advancements in Section~\ref{Disc}. Finally, we conclude with a summary in Section~\ref{Summ}.
\section{Observational Data}
\label{Data}

The observed flare event was occurred in AR NOAA 11166 on March 9, 2011 located in north-west (N23$^{\circ}$W$11^{\circ}$) part of the solar disk. The observations of the event (from 20:00\,UT on 9 March to 02:00\,UT on 10 March) are well covered by \textit{Helioseismic Magnetic Imager} (HMI; \citealt{schou2012}) in 6173\,\AA~providing photospheric vector magnetic field information at 0.5 arcsec/pixel resolution, and \textit{Atmospheric Imaging Assembly} (AIA; \citealt{lemen2012}) in 10 wavelengths providing solar atmospheric imaging information at 0.6 arcsec/pixel resolution. 

The AIA images are processed using standard procedures in SolarSoft and reduced to one minute cadence by integrating over five successive images in order to remove noise. The HMI stokes parameters are derived from filtergrams averaged over a 12 minute interval, and then they are inverted with the help of a Milne-Eddington atmospheric model, using the Very Fast Inversion of Stokes Vector (VFISV; \citealt{borrero2011}) algorithm to yield all three components of the photospheric magnetic field vector. The inherent $180^{o}$ azimuth ambiguity in the transverse field is removed using an improved version of the ``Minimum Energy'' algorithm \citep{metcalf1994, metcalf2006, leka2009}. After correcting for projection effects \citep{venkat1989, gary1990} the data were re-mapped and transformed to heliographic coordinates from spherical coordinates using Lambert cylindrical equal-area projection \citep{calabretta2002}. After all these procedures, the mean value of the estimated errors range up to 25, 50\,G in the line-of-sight and transverse components, respectively. Therefore, we consider only pixels with values above the threshold ($|B_t|>150\,G$ and $|B_z|>50$\,G). Detailed information about retrieving the final vector field products from level zero filtergrams are available in \citet{hoeksema2013}.

Complementary chromospheric observations from GONG site at Big-Bear Solar observatory and hard X-ray information from \textit{Reuven Ramaty High-Energy Solar Spectroscopic Imager} (RHESSI; \citealt{lin2002}), are also used in this study. All  maps are aligned with the magnetograms and projected onto the disk center.

\section{Coronal Magnetic Field Extrapolation and QSL calculation}
\label{Methods}
In order to study the 3D-magnetic field evolution above the AR in association with the flare, the photospheric vector magnetic fields are extrapolated with the help of a Non-Linear Force-Free Field algorithm (NLFFF; \citealt{wiegelmann2010,wiegelmann2012}). The algorithm minimizes the functional

\begin{multline}
L =\int\limits_{V}{\left( w\frac{|(\nabla \times \mathbf{B})\times \mathbf{B}{{|}^{2}}}{{{B}^{2}}}+w|\nabla \bullet \mathbf{B}{{|}^{2}} \right)}dV+ \\
    \nu\int\limits_{S}{\left( \mathbf{B}-{{\mathbf{B}}_{obs}} \right)}\cdot W \cdot \left(\mathbf{B}-{{\mathbf{B}}_{obs}} \right)dS
\label{Eq_Func}
\end{multline}

The first integral contains quadratic forms of the force-free and solenoidal  conditions. $w$ is a weighting function towards the lateral and top boundaries. In the original approach by \citep{wiegelmann2004} the functional $L$ is iteratively minimized with boundary conditions derived from measurements of the photospheric magnetic field vector. The newly added surface integral term takes into account measurement errors and allows a slow injection (controlled by the Lagrangian multiplier $\nu$) of the boundary data \citep[see][for details]{wiegelmann2010}. {\bf W}(x, y) is a diagonal matrix, which is chosen inversely proportional to the transverse magnetic field strength.

While iteration of NLFFF algorithm, the following quantities are monitored 
\begin{align}
{{L}_{1}}&=\int\limits_{V}{\frac{|(\nabla \times \mathbf{B})\times \mathbf{B}{{|}^{2}}}{{{\text{B}}^{2}}}}\text{dV}; \nonumber \\
{{L}_{2}}&=\int\limits_{V}{|(\nabla \bullet \mathbf{B}{{|}^{2}}}\text{dV}; \nonumber \\
{{\sigma }_{j}}&=\left( \sum\limits_{i}{\frac{|\text{J}\times {{\text{B}}_{i}}|}{{{\text{B}}_{i}}}} \right)/\sum\limits_{i}{{{\text{J}}_{i}}}
\label{Eq_Diag}
\end{align}

\noindent Where $L_1$ and $L_2$ corresponds to the first and second terms in Equation~\ref{Eq_Func}, respectively. These integrals consider only the inner $236\times236\times188$ physical region (where the weighting functions are unity) and $\sigma_j$ corresponds to the sine of the current weighted average angle between magnetic field and electric current density. During the iteration these terms decrease and force the electric current parallel to the magnetic field (force-free). For our data sets $\sigma_j$ becomes finally about $15^{\circ}$.

The initial potential fields (PF) are computed with the help of a Green's function method \citep{sakurai1989} and require only the normal component of the photospheric magnetic field. The PF solutions are used as initial state and also to prescribe the top and side boundaries for the NLFFF algorithm. Our computations are performed in a Cartesian box with a grid of $300\times300\times160$, which corresponds to physical dimensions of $219\times219\times117$\,Mm$^3$ encompassing the AR on the sun. As the vector magnetograms of this AR satisfy the force-free consistency conditions approximately (flux imbalance less than 1\%,  force and torque terms are about 5\%), preprocessing of the data was not necessary \citep{wiegelmann2006, wiegelmann2012}.

After constructing 3D coronal fields, the change in magnetic fieldline linkage is measured by the strength of QSLs defined as squashing degree Q \citep{titov2002, titov2007}. Q describes the fieldline mapping gradients and larger values of Q corresponds to the cross section of quasi-separatrix layers in any plane. It is computed by tracing consecutive two fieldlines with footpoints at extremely small distance and measuring the distance between the respective conjugate footpoints as given by 

\begin{equation}
Q=\frac{{{\sum\limits_{i,j=1}^{2}{\left( \frac{\partial {{X}_{i}}}{\partial {{x}_{j}}} \right)}}^{2}}}{|{{B}_{z,0}}/{{B}_{z,1}}|}
\label{Eq_SqFac}
\end{equation}
\noindent where $X_i$(i=1,2) is the coordinates of conjugate foot point in Cartesian system, $B_{\rm z,0}$ and $B_{\rm z,1}$ are vertical field components at start and end footpoints of a fieldline. For dense points of Q, the calculations are performed in a grid of resolution increased by eight times that of extrapolation. We use an iterative scheme as described in earlier studies \citep{aulanier2005,savcheva2012,xudong2013}. The QSL is defined by $Q>>2$, and the value Q=2 is the lowest possible value. The maximum value of Q found in our computation is $\sim$$10^{13}$ but much larger values are possible at still higher resolution.   

\begin{figure*}[!htbp]
\centering
\includegraphics[width=0.99\textwidth,clip=]{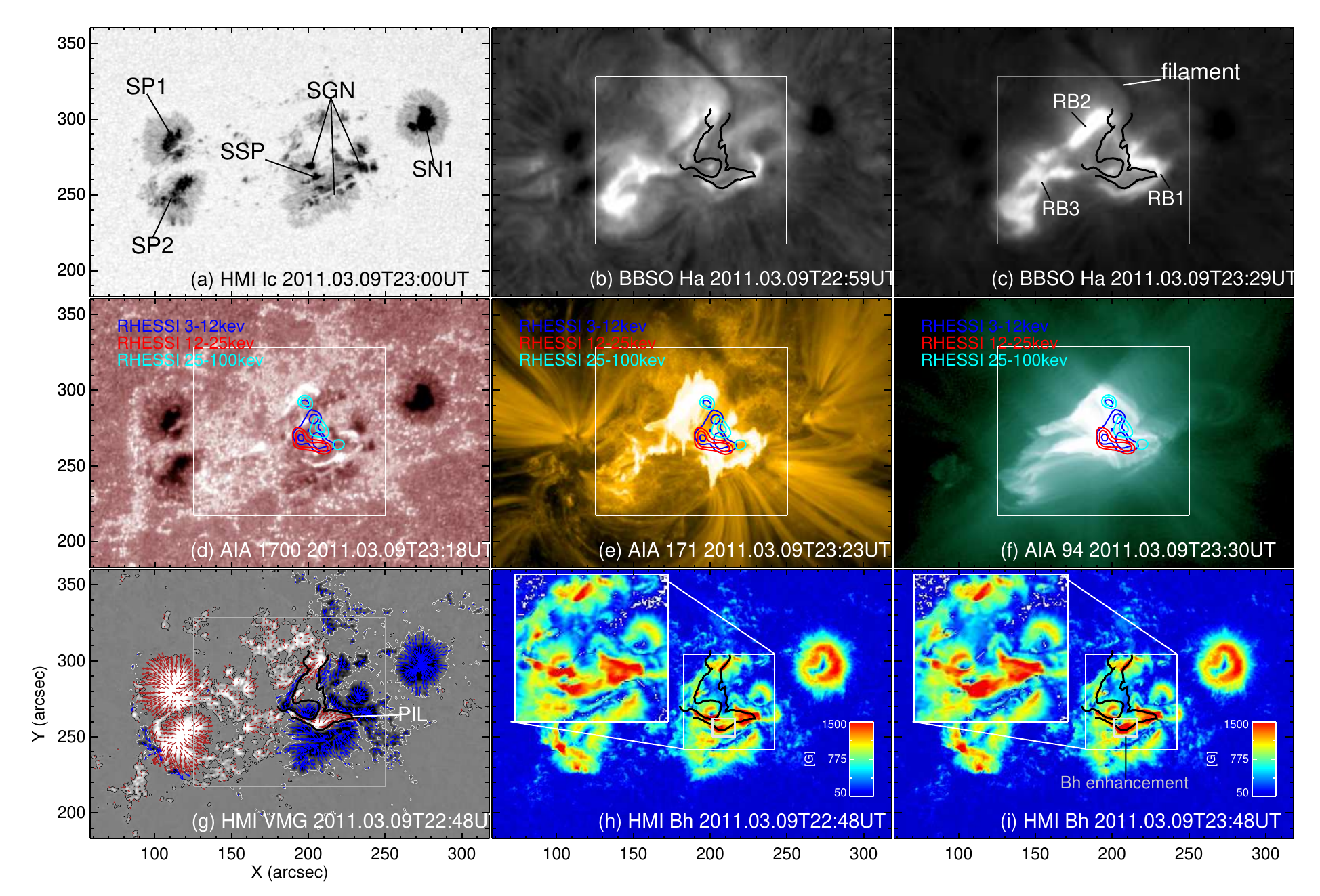}
\caption{Different observations of AR 11166 around the time of the X1.5 flare (a) Continuum intensity showing prominent sunspots, as marked, (b-c) Chromospheric H$\alpha$ images with the apparent quasi-elliptical shape ribbon RB1 and the remote ribbons RB2, RB3. Black traces indicate the  existence of a filament channel originating from ribbon RB1, and black curves refer to the magnetic polarity inversion lines (d) Similar signatures of ribbons are also visible in UV 1700\,\AA~images, (e) AIA 171\,\AA~image showing the intense brightness along the flare ribbons. The blue, red, cyan colored contours indicate RHESSI HXR sources reconstructed with the ``clean'' algorithm, plotted at 80, 90\% levels of maximum in the 6-12, 12-25 and 25-100\,kev bands, (f) Another image showing the connection between the coronal energy release locations with chromospheric ribbons just after the flare maximum, (g) HMI vector magnetogram (VMG) showing the horizontal field vectors (blue/red arrows represent direction and length proportional to the magnitude) over plotted onto a $B_z$ map with contours (black and white) at $\pm$150\,G. Notice the sunspot SSP is associated with positive polarity flux which is surrounded by the sunspot group SGN, which is associated with negative flux. In the flaring location, the PIL (thick curve), separating positive and negative polarity regions, almost follows the flare ribbon geometry. The flaring site is highlighted by a rectangular box in panels (a)-(f), and will be referred as region-of-interest (ROI) for further studies, (h-i) Maps of the horizontal field ($B_h$) distribution before and after the flare, plotted with colour scale (20-1500\,G). The flaring location is zoomed and shown in the inset, indicating the field enhancement at the localised position marked by an arrow.}\label{Fig_Mos}
\end{figure*}
\section{Flare Evolution, Morphology and Magnetic Field Properties}
\label{FlareMorphEvol}
The AR 11166 appeared on the northern part of the solar disk on March 3. Although it was already well emerged, it continued to evolve with further flux emergence. During disk transit, the region was very dynamic and harboured several flares and CMEs. The AR and its long term activity, caused by flux motions and helicity injection was thoroughly investigated  in \citet{vemareddy2012a}. The second X-class flare in the 24 solar cycle X1.5 occurred in this AR on March 9, at 23:13\,UT, and we study its structural evolution here. In this context, we focus our investigation on 6 hours of observations around the time of the flare.   

Figure~\ref{Fig_Mos} shows the observations of the flare event in different wavelength-passbands. As visible in the HMI continuum images (panel \ref{Fig_Mos}(a)), the AR consists with sunspots of variable size, which have well developed umbra and penumbra. Two of the sunspots are located on the East (SP1, SP2) and one on the West (SN1) are large in size. Smaller and fragmented spots are grouped around the center part of the AR. The chromospheric \Ha~images show the presence of a dark filament with one leg originating from the central portion of the sunspot group and the second leg is anchored near sunspot SP1. Signatures of bright ribbon emissions are observed from 20:00\,UT onwards. They appear only during the flare, however. 

From the GOES soft X-ray profile, the flare initiation occurred at 23:13\,UT, having peak phase emission in hard X-rays at 23:21\,UT. With progressive reconnection in the corona, emissions eventually form ribbons that start appearing from 23:15\,UT onward and exhibits peak emission about the peak time of the flare (23:23\,UT). The light curves of \Ha~and UV 1700\,\AA~exhibits their maximum flare emission almost at the same time. The EUV emission peak in 171, 94\,\AA~channels, is observed, however, only later at 23:29\,UT and is associated with post flare loops, which evolve further for another half hour.

The main flare ribbon RB1 has a ``quasi-circular'' shape which formed within the sunspot group. RB1 seems to have close connection to the coronal reconnection site. In addition, two ribbons (RB2 and RB3) formed remotely from RB1 more or less simultaneously. An important aspect of this event is that it has no evident ribbon motions, but it expands within one minute during the impulsive phase. It is assumed that static ribbons are associated with stationary coronal magnetic reconnection. Motions during the progressive reconnection phases are seen, however, for events close to the limb (e.g., \citealt{bhuwan2009}).

From the photospheric  magnetic field distribution, we understand that the ribbons form along the polarity inversion line (PIL), which separates the two opposite polarities (Figure~\ref{Fig_Mos}(b-c)). This central portion has a peculiar magnetic field distribution: the positive flux of the small sunspot SSP is embedded in a negative flux region associated to the sunspot group SGN (panels {a) and (g)  in Figure~\ref{Fig_Mos}). This flux distribution forms a nearly closed contour of the PIL (thick curve in panel (g)). Such configurations are prone to contain topological structures like null points and are potential locations where current sheets may form \citep{demoulin1993,demoulin1997}. Despite this complex flux distribution, the overall magnetic configuration of the AR is bipolar (with imbalance of fluxes less than 1\%) with a dominantly negative polarity in the West and a positive one in the East.

In coronal AIA 171\,\AA~images, we see coronal loops connecting the major sunspots of opposite polarity. The magnetic connectivity in the flaring region was somewhat difficult to identify before the event, but in later times the formation of post flare loops are well visible. During the peak phase, the emission along the PIL coincides spatially in all AIA channels and with the \Ha~ribbons. The coronal loops filled with plasma are regarded as tracers of the magnetic fieldlines and help to determine the structure of the observed magnetic flux distribution. The fieldline connectivity is clearly visible in the central flaring part of the AR. This gives us some clues about the relation of the coronal field structure and the ribbons. From observations of hot coronal loops in 94\,\AA~we conclude that the coronal fieldlines have their footprints located within the observed emission of ribbons.

To relate these emissions along the ribbon with the reconnection site in the corona, hard X-ray sources are overlaid [contours of 80, 90\% of maximum] on EUV iamges (Figure~\ref{Fig_Mos}(d-f)). The hard x-ray (HXR) images are reconstructed from ``clean'' algorithm \citep{hogbom1974} at one minute integration time. The peak HXR emission coincides with that of \Ha~and UV, implying their spatial and temporal association during the flare. Indeed, the HXR sources in 6-12, 25-100\,kev bands fall in between RB1 and RB2. We co-aligne the images to an accuracy level of 2-3 arcsec by using magnetic field contours on different instrument images to compute the corresponding shifts. However, due to problems with projection effects, an ambiguity persists to differentiate between HXR sources from footpoint and loop top sources. These observations imply that RB1 and RB2 are formed by direct impingement of particles from the reconnection site in the corona. 

For investigating the flare emission in $H\alpha$ in more detail, we plot light curves from RB1, RB2 and RB3 (not shown in this paper). The curves show that emission from RB1 and RB2 simultaneously, the emission from RB3 is delayed by about one minute when compared to earlier one. This delayed emission from RB3 could be likely related to its remote presence from flaring site. All of these observational facts reflect the scenario of the classical flare model: coronal plasma particles become accelerated in the corona by magnetic reconnection and traverses downwards along the fieldlines and thereby bombarding the low coronal dense material, which consequently leads to the formation of ribbons with bright emission.

\begin{figure}[!ht]
\centering
\includegraphics[width=.49\textwidth,clip=]{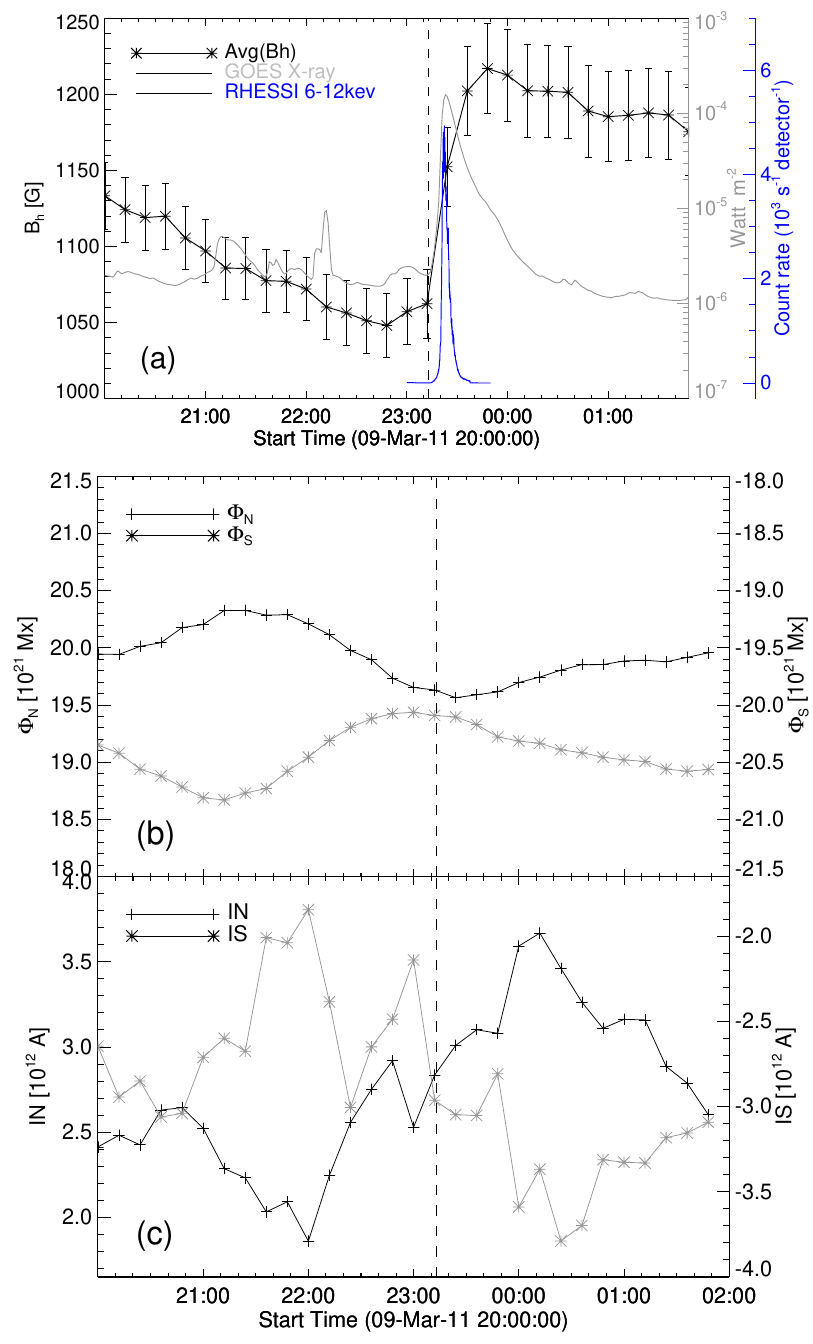}
\caption{(a) Time profile of the average horizontal magnetic field ($B_h$) in the localized rectangular region shown in Figure~\ref{Fig_Mos}(h). GOES soft X-ray flux (grey) and RHESSI hard X-ray flux (6-12\,kev band in blue) are also plotted with y-axis scale (in blue) on the right side of the plot. The error bars refer to 3-sigma levels of standard deviation, (b-c) Time profiles of integrated line-of-sight magnetic flux $\Phi$ and net vertical current (I) at North (N) and South (S) polarities in the AR. The vertical dashed line in all panels indicates the start time of the X1.5 flare, as observed with.}\label{Fig_Wang}
\end{figure}

The vector magnetogram, horizontal field vectors are overplotted onto the vertical magnetic field maps panel~\ref{Fig_Mos}(g)) illuminates the sheared field distribution along the PIL. In the presence of continued slow flux motions as the AR evolves, stress can build up and store energy in the magnetic configuration. Please note that the horizontal field vectors are almost aligned to the PIL and reflect the presence of such stressed fieldlines. All these observations indicate favourable conditions for flare (e.g., \citealt{vemareddy2012b}).

Several studies \citep{sudol2005, petrie2010, wangh2010} gave evidences that flares may change the magnetic field configuration significantly in a persistent and permanent manner. The persistence of the observed field changes imply that they are not artifacts of changes in the photospheric plasma parameters \citep{maurya2012} during the flare, and the temporal and spatial coincidences between flare emission and field changes suggest that field changes are linked to the flare. To check the presence of such flare associated field changes, we have scrutinized horizontal field maps ($B_h=\sqrt{B_x^2+B_y^2}$) as plotted in Figure~\ref{Fig_Mos}(h-i) before (22:12\,UT) and after (23:48\,UT) the flare. In the insets, large horizontal fields (up to 1500\,G) have been distributed along the PIL. Their prominent change spreading in a large area is identified, however, only at a localized region (small rectangular box). From this small region, the average $B_h$ is evaluated and plotted with respect to time in Figure~\ref{Fig_Wang}(a). Obviously, the evolution of horizontal field at this location during the flare delineates that they have undergone a permanent change with a net average increase of about 180\,G (see panels~\ref{Fig_Mos}(g-h)). These observations are consistent with recent reports of collapsing fields appearing as increased horizontal field after reconnection \citep{wangs2012}.

Similarly, the net vertical current ($\text{I}=\sum\limits_{N\,pixels}{{{\text{J}}_{\text{z}}}}$) and net line-of-sight flux ($\Phi =\sum\limits_{N\,pixels}{{{\text{B}}_{\text{z}}}}$) in the AR is computed at North (N) and South (S) polarity, and we show their temporal evolution in Figure~\ref{Fig_Wang}(b-c), respectively. The flux in both N- and S-polarity varies by $0.5\times10^{21}$\,Mx over the time interval of 6\,h indicating that the variations are likely part of the general evolution but not related to significant flux emergence or cancellation. Since positive (negative) flux is having positive (negative) signed current, it is likely that the AR magnetic field is right helically twisted \citep{wangj2004} with positive helicity, which is consistent with the sign of helicity injection rate derived from flux motions \citep{vemareddy2012a}. This interpretation is in turn supported by the sign of average alpha given by
\begin{equation*}
\alpha_{\rm av} = \frac{\sum J_z(x,y){\rm sign}[B_z(x,y)]}{\sum |B_z|}=(1.27\pm0.19)\times10^{-8} m^{-1}
\end{equation*}
computed at photospheric boundary, which implies an average twist in the AR. The net vertical current, as evaluated from both polarities increased from 22:00\,UT towards their peak values at the end of the day. The net current changed by a similar amount of about $1.8\times10^{12}$\,A in both signed fluxes, which implies that they are likely generated around the PIL by shearing motions. This significant increase of net currents is usually a precursor for flaring activity \citep{schrijver2008,ravindra2011}

\begin{figure*}[!ht]
\centering
\includegraphics[width=0.99\textwidth,clip=]{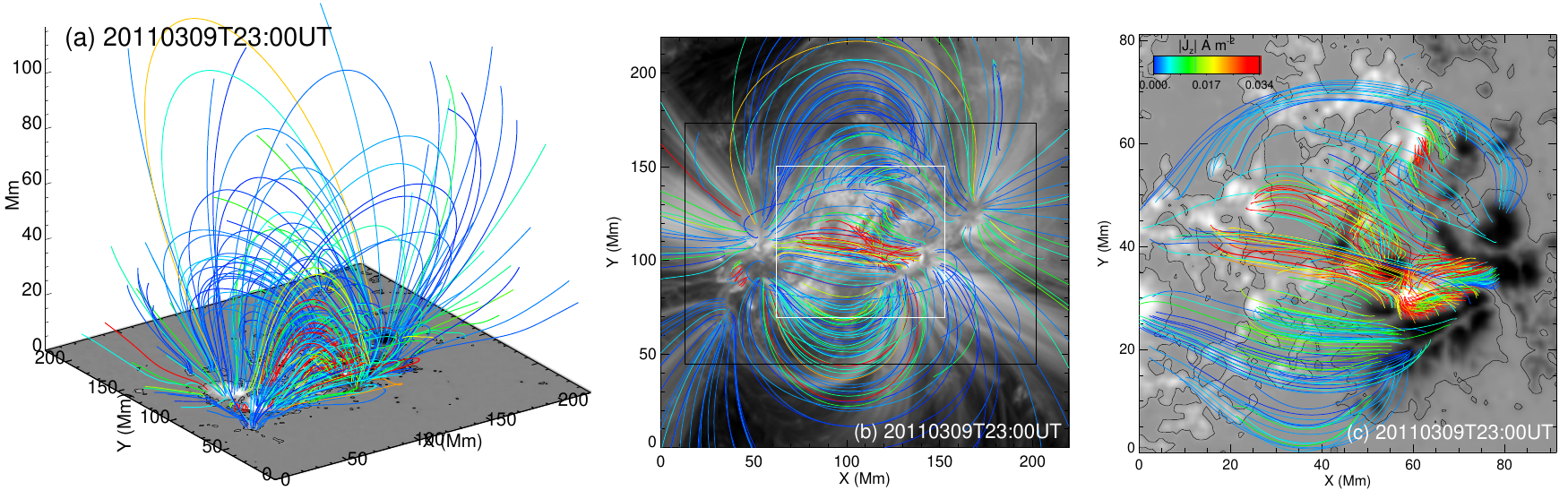}
\caption{Coronal magnetic field structure (derived from extrapolations) above AR 11166 on March 9, 2011 at around 23:00\,UT. (a) The fieldlines are overplotted onto the normal photospheric field. The box contains the full computation domain of $300\times300\times160$ grid cells (1 grid cell$=0.725$\,Mm) displayed in 3D. (b) Fieldlines projected onto an AIA 171\,\AA~passband image. Note that the fieldlines resemble the plasma tracers well, thereby indicating the validity of the NLFFF coronal magnetic field model. The black rectangular box represents the field-of-view (FOV) of the AR displayed in Figure~\ref{Fig_Mos} and the white one indicates the flaring region including the extended ribbons, hereafter refer as region of interest (ROI), (c) the ROI with fieldlines that anchored within the FOV. In all (a-c) panels, the fieldlines are color coded according to the vertical current at the lower boundary ($J_z[z=0]$) as shown with colorbar in (c). Generally, most red color fieldlines located around the positive flux embedded in surrounding negative flux, i.e., along the PIL.}\label{Fig_Topo1}
\end{figure*}

\section{Coronal Magnetic Field Structure and Fan-Spine Topology}
\label{MagStrTopo}
In order to derive the magnetic connectivity of the observed
boundary distribution of the fields, we constructed the 3D-field above the AR using potential and NLFFF approximations as discussed in Section~\ref{Methods}. The force freeness of the reconstructed NLFFF solutions is measured by weighted average angle ($\sigma_j$) between field and current which is less than $15^{o}$ and the integrals in Equation~\ref{Eq_Diag} are about $L_1\le$2.7 and $L_2\le$1.3 in all time frames. In principle, these terms must vanish for exact force-free equilibria, but due to noise and measurement errors in the boundary data, small residual forces remain. The solutions are still valid for further analysis, however.

Figure~\ref{Fig_Topo1} shows fieldlines (colour coded with the vertical currents at their footpoints) at 23:00\,UT. From the oblique view in panel~\ref{Fig_Topo1}(a), the fieldlines connecting the major sunspots are overlying (up to the heights of 110\,Mm) to those (below 40\,Mm) connecting the fluxes at the middle part of the AR. In panel~\ref{Fig_Topo1}(b), the same fieldlines are plotted onto an 171\,\AA~channel image for comparing them with coronal plasma loops as proxies of magnetic fieldlines. Plasma structures from major sunspots resemble those with less current flowing fieldlines. This indicates that these overlying long loops are almost current free. However, compact, stressed locations are generally observed at flaring sites (white rectangular region shown in panel~\ref{Fig_Topo1}(c)). These complicated structures are difficult to model even with the NLFFF approximation. As conjectured earlier, sheared field vectors along the PIL around SSP yields stressed (in red-) fieldlines with significant currents along them. It is worth to mention that this is the only location that differs from the rest when compared to that of PF based magnetic structures. These elevated structures indicate a non-potential, higher energy state in the flaring region. Those fieldlines which are somewhat away from the PIL in the surrounding negative flux region are connecting to remote positive flux regions. They are passing over the isolated positive polarity SSP and covering it with a dome-liked structure. Such configurations are possible locations to form null points with a specific topology relevant for reconnection. In total, the NLFFF model reproduced field structures that resemble coronal loops at large scales and at small, compact locations the structure is stressed as expected from the sheared boundary vectors (e.g., \citealt{vemareddy2013}). 

\begin{figure*}[!htp]
\centering
\includegraphics[width=0.99\textwidth,clip=]{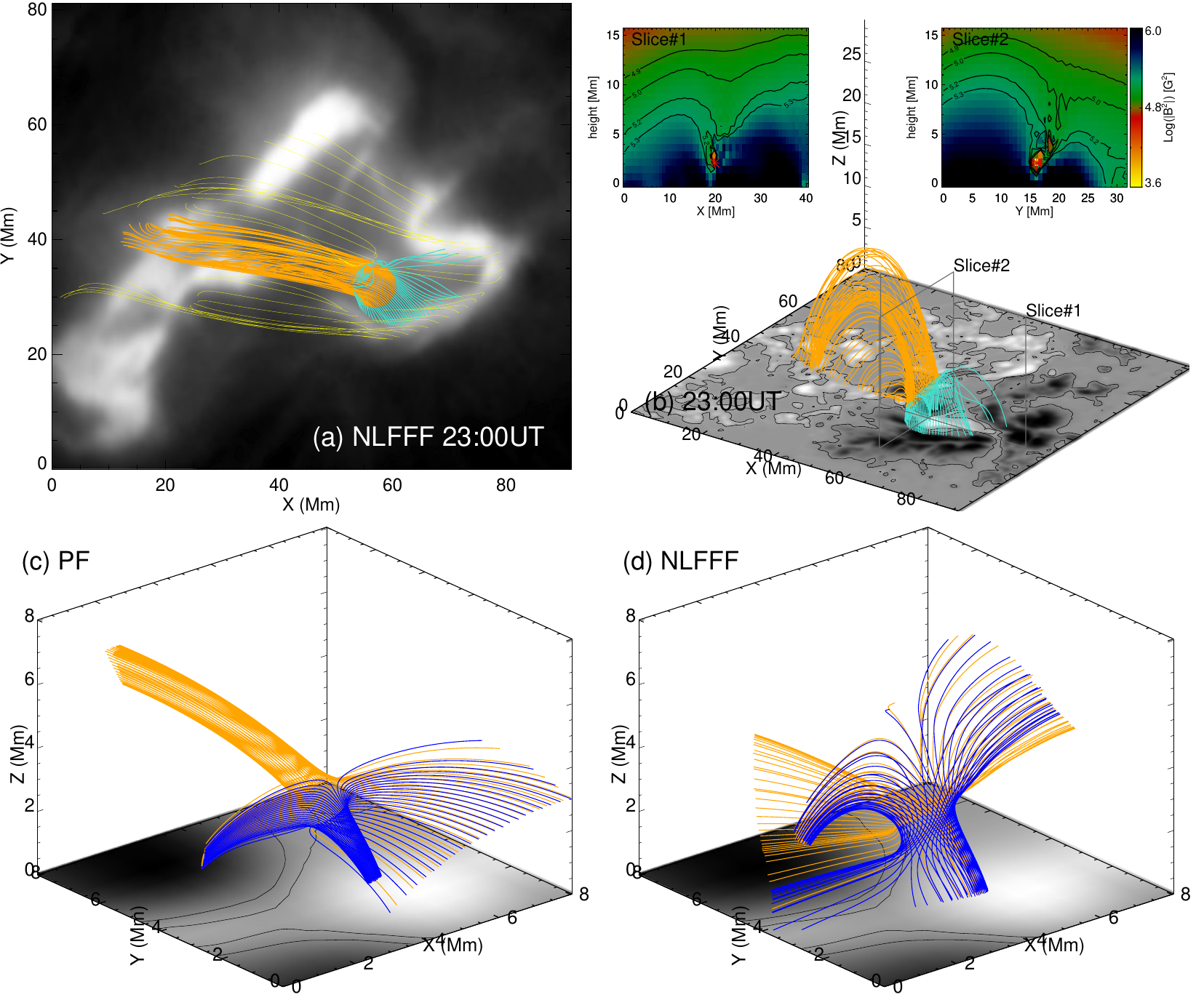}
\caption{Fan-Spine topology in AR 11166 (a) Top view of topological skeleton with dome-shaped fan fieldlines (cyan) landing along ribbon and spine fieldlines (orange) connecting remote ribbon RB3. Note that the fieldlines from about outer edge of the ribbon RB1 are deployed in yellow, showing its obvious connectivity with extended remote ribbons RB2 and RB3. (b) Same structure plotted on $B_z$ boundary map depicted in 3D view. The field $|\textbf{B}|$ is extracted in the cross-section planes of slice$\#1$, slice$\#2$ containing the null point and displayed in the insets. Note the null location approximately at 2.11\,Mm in both inset plots. (c) Close-up view of the magnetic structure in PF model around the null in the volume of 8\,Mm side of cube. Bottom plane is $B_z$ with contours at $\pm150$\,G. Fieldlines below (above) the fan plane are colored in blue (orange) for clarity. Evidently, it delineates fan-spine topology with fan fieldlines constructing $\Sigma$-surface, inner and outer spine fieldlines as $\gamma$-line. (d) Same as panel (c) in NLFFF model. The presence of a fan-spine topology is obvious. The spiral nature of the fieldlines (in blue) indicate the presence of current and twist  around the null.}\label{Fig_Topo2}
\end{figure*}

Finding the null point and topological properties in a noisy reconstructed field structure is a difficult task. The procedure described in \citep{haynes2007} is adopted to find the exact position of the null point. Basically, it involves scanning for the null to locate possible grid cell and finding the precise position using a tri-linear interpolation with the help of an iterative Newton-Raphson scheme within the grid cell. We found null points in every frame of the PF model from 20:00\,UT to 23:36\,UT. In the NLFFF model, however, the algorithm did not find any null in a few frames, probably because of noise and computational errors. In the NLFFF field, the location of the null is (78.04, 43.55, 3.76) at 23:00\,UT frame in the ROI. In all other frames, the null point is observed about this grid cell at a height of about 2.11\,Mm (1 grid cell = 0.725\,Mm) and above. The reason to find no null after 23:36\,UT in either model would likely related to a structural change of the magnetic configuration by reconnection during the flare.

Further, topological properties of null can be obtained from eigenvalues and eigenvectors of the Jacobian matrix  $\delta B={{\nabla }_{j}}{{B}_{i}}=\partial {{B}_{i}}/\partial {{x}_{j}}$ \citep{lau1990, priest1996} in the vicinity of the null. For instance, the eigenvectors 
\begin{eqnarray*}
{{\hat{e}}_{0}}={{[-0.4181, 0.2368, 0.8770]}^{\rm T}} \\
{{\hat{e}}_{1}}={{[0.6026,-0.7924,0.0948]}^{\rm T}} \\
{\hat{e}}_{2}={{[-0.0039, 0.9791, 0.2031]}^{\rm T}} 
\end{eqnarray*}
corresponding to eigenvalues (-222.15, 105.87, 147.47) are evaluated for the $\delta B$ matrix at 23:00\,UT frame. As the determinant (product of eigenvalues) of this matrix is negative, the type of null is B or positive \citep{parnell1996}. The nulls found at different time frames from NLFFF solution are positive (B-type) whereas in the PF solution, the found nulls are all negative (A-type or determinant of Jacobian matrix is greater than zero).

With $\delta B$, Lorentz force (F) and electric current ({\bf J}) can also be estimated locally. The trace of $\delta B$ must vanish. This is not the case, however, because of numerical errors and linearization. The average relative error \citep{xiao2006} for $\delta B$ is estimated with the ratio given as, $|\nabla \bullet \mathbf{B}|/|\nabla \times \mathbf{B}|\approx 10\%$. At this level of precision, the NLFFF approximation is presumed to be a suitable model reproducing and identifying reliable orientation of the fan-spine structure. Boundary data with less noise might lead to more accurate solutions in the form of the averaged angle ($\sim$15$^{\circ}$) between current and field in NLFFF solutions.

Two of the eigenvectors (with same sign of the eigenvalues) define the fan surface and the third one specifies the spine direction. With knowledge of the fan plane orientation, fieldlines away from the null can be traced in a circle of points on either side of the fan-plane to visualize the local structure of the null. As an exemplary case, these fieldlines are plotted on an \Ha~image in Figure~\ref{Fig_Topo2}(a). The inner (cyan) and outer (orange) fieldlines meet in a circle of points that define the fan plane which is oriented at a tilt angle. The fieldlines below the fan plane (cyan) are diverging to connect with the photosphere exactly along the ribbon RB1 and those above (orange) are converging to connect at RB3.

In panel~\ref{Fig_Topo2}(b), the same structure is illustrated in oblique 3D view, over plotted onto the photospheric $B_z$ field. The inner fieldlines (cyan) form a dome-like structure, which completely covers the isolated positive polarity spot SSP. This fan dome divides the volume into two distinct connectivities. To prove the existence of a null point, we extract the field strength $|{\bf B}|^2$ distribution in two perpendicular planes of slice\#1 and slice\#2. They are shown in insets with logarithmic scale in the same panel. From the field distribution in these two slice planes, it is obvious that the residual field (yellow) is embedded within the strong fields of the AR and assures the correct identification of the null point. 

In addition, we also traced fieldlines that have footprints lying along the outer edge of ribbon RB1 and are shown in the same panel (in yellow). They connect to the remote ribbons RB2 and RB3 demonstrating their topological relation. The connectivity with the south-east portion of RB3 from RB1 have no exact topological relevance, but the connection with the far away negative flux from RB1 is visible in Figure~\ref{Fig_Topo1}(c). These observations provide straightforward evidence that the identified coronal null point in the magnetic field structure confirms to be associated with a fan-spine topology, which is of specific importance for favouring reconnection.

In panel~\ref{Fig_Topo2}(c), the fan-spine structure in the PF solution is obtained in a small volume of $8\times8\times8$\,Mm$^3$ centered around the null. At this small length scale, the noise in the numerically extrapolated field did not allow to directly compute the magnetic topology, until we smooth the field with a boxcar of three grid cells. As before, fieldlines in a circle of radius (0.7\,Mm) from the null are traced in up and down side of the fan plane. Not surprisingly, the null point contains two types of isolated fieldlines called the spine lines. One is below the fan plane (blue) touching the middle part of isolated sunspot and another is above the fan plane (orange) and extends towards RB3. In the fan plane, both of the fieldlines (blue and orange) diverge and manifest a separatrix surface in the form of a dome shaped structure intersecting with the photosphere along the inner edge of the ribbon. This local field structure around the null is reminiscent of theoretically predicted topology by \citet{lau1990} (see their Figure 2) with a $\gamma$-line along the spine and a $\Sigma$-surface as fan. Owing to the type of the null which is negative (A-type), the fieldlines in the fan plane are directed away from the null \citep{parnell1996}. Please note that the spine eigenvector is almost normal (angle deviation of $2^o$) to the fan plane, similar as expected for potential field nulls.

Similarly, for the NLFFF solution (panel~\ref{Fig_Topo2}d), the existence of a fan-spine topology is obvious, but it is structurally different from the characteristics of a PF magnetic null. The eigenvectors constructing the fan plane subtend a $139^{\circ}$ angle and the normal has about a $24^{\circ}$ angle with the spine axis, leading to a tilted orientation to the fan plane. An electric current (${\bf J}.{{\hat{e}}_{0}} \neq 0$,${\bf J}.{{\hat{e}}_{1}} \neq 0$,${\bf J}.{{\hat{e}}_{2}} \neq 0$) flows along the eigenvector directions which is consistent with the nature of the non-potential null and a non-perpendicular orientation of the fan plane with the spine axis. Moreover, because the type of this null is positive (B-type), the fieldlines must direct towards the null and we can notice from the figure that they are spirally approaching the null. This spiral nature of the fieldlines around the null delineates twist and currents forming a complex topological structure which is entirely different to that of potential case.

Thus, we identify a magnetic null point in the coronal magnetic structure that is associated with a specific fan-spine topology. The intersection of fan dome with the photosphere along the main flaring ribbon provides potential evidence for the topological relevance of coronal reconnection location and ribbon emission.

\begin{figure*}[!ht]
\centering
\includegraphics[width=.99\textwidth,clip=]{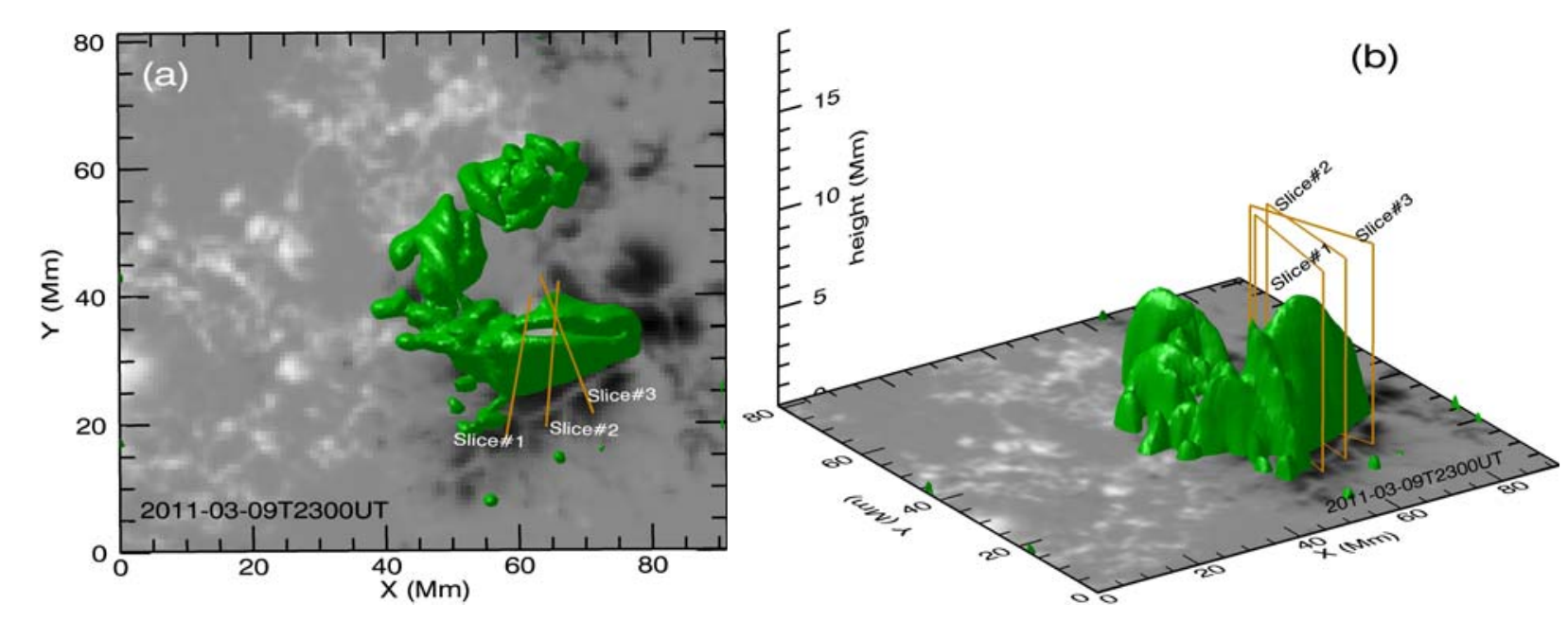}
\caption{Iso-surface (level 0.015Am$^{-2}$) of coronal current distribution $|\textbf{J}|$ projected onto the photospheric $B_z$ magnetogram (a) in plane projected 2D view (b) perspective 3D view. Three planes of slice\#1, slice\#2, slice\#3 [orange rectangular planes], the cross-section of the domed structure of the fan-spine topology are considered to extract their current distribution (Figure~\ref{Fig_Slj}). Obviously, the current is concentrated at the localized fan separatrix (see Figure~\ref{Fig_Topo2}(a)) above the PIL in the corona, up to 8\,Mm and intersecting along the ribbons at the main flaring location.}\label{Fig_Isurf}
\end{figure*}

\section{QSLs, Current Distribution and Reconnection}
\label{RibCurrRec}

Using the three-components of the magnetic field, we compute the electric current distribution $|\textbf{J}|$ above the flaring region and study its evolution. In Figure~\ref{Fig_Isurf}, the iso-surface of $|\textbf{J}|$ at 0.015\,Am$^{-2}$ level is rendered and plotted onto the $B_z$ magnetogram in top and perspective views. The chosen level of iso-surface indicates a strong current distribution. By comparing these coronal current with the locations of the \Ha~flare ribbons, we understand that ribbon RB1 is directly associated with a coronal current layer as a domed structure. The intersection of this current layer at the lower boundary almost traces the path of RB1. Furthermore, its volume distribution is concentrated at the 3D separatrix surface of the domed structure, covering SSP as seen in Figure~\ref{Fig_Topo2}(b). This indicates that the separatrix layers constituted by the domed fieldlines (Figure~\ref{Fig_Topo2}b) are natural locations for current layers, as found in several theoretical studies \citep{demoulin1996, demoulin1997}. As the ribbon RB1 follows the path of the PIL, which is surrounded by a sheared horizontal field distribution (Figure~\ref{Fig_Mos}, panels g-i), the currents are generated by twisted structures along the PIL extending to the corona and forming an almost closed surface. As no flux emergence appeared, the assumed mechanism to build these coronal currents is the motion of flux. At higher values of the current, the iso-surface can still be thinner and lower in height. Two similar locations of the coronal current locations are identified along ribbon RB2. However, above the ribbon RB3, the current is weakly distributed compared to these locations, so that its iso-surface cannot reach prominent levels.

In the successive stages of the quasi-static evolution, the stress in the photospheric fields develops possibly by slow flux motions, which in turn build up coronal currents. These stressed fieldlines about the closed PIL, construct a domed coronal current distribution. It is important to note that the null is located within this dome structure. When reconnection commences at this coronal null, the currents at the associated separatrices will dissipate and power the flare. In order to capture the effect of dissipation, we have extracted $|\textbf{J}|$ in the planes of arbitrarily chosen slices\#1,2,3 cross sectioning the dome-structure as depicted in Figure~\ref{Fig_Isurf} (orange planes). The obtained current distribution in these slice planes from the time frames before (22:00\,UT), during (23:12\,UT) and after (23:48\,UT) the flare are plotted in Figure~\ref{Fig_Slj}. The height of the current distribution extends to 8\,Mm (the contour level 0.01\,Am$^{-2}$ is marked) in the corona. Note that slice\#1 is almost in the plane containing the coronal null and crosses the localized region involving photospheric field enhancement, as noted in the previous section.

This coronal current distribution in slice\#1 evolves towards a constricted shape at 23:12\,UT, thereby forming a cusp shaped distribution appearing at lower height at 23:48\,UT. A similar evolution is seen mildly in the current distribution of slices\#2, \#3. As the coronal current evolution is affected more in slice\#1, having a magnetic null point, it confirms further that reconnection occurs in the corona in this plane, but not in the planes across ribbon RB1, which do not contain nulls. Note that reconnection of magnetic fields is a dynamical effect and occurs in thin current sheets which are not resolvable by present observations. The coronal current shown here, has been computed from the curl of the slow evolving coronal fields.
 
It is highly difficult to trace the same single fieldline in every frame exactly to trace its connectivity change during the reconfiguration for several reasons. Because the reorganization of the fields happens in the corona, the photospheric fields would respond to such an effect in the line tied scenario. However, this effect of coronal reconfiguration on the photosphere is small, localized and sometimes not present at all. One can better investigate such changes, by computing a number of collected fieldlines from a localized region. The coronal field evolution in the domed-structure is shown in the panels of the last column in Figure~\ref{Fig_Slj}. Fieldlines above the PIL at two different heights (1\,Mm and 4\,Mm) are traced and plotted. Their configuration indicates a sheared nature and provides evidence for coronal currents in the form of ridges (obtained in slice planes across the dome) as shown in the other panels of the same Figure. The overlying fieldlines (orange) connect to distant fluxes around the PIL and some of them are open to converge with the spine from the null location. In successive time frames, these open overlying fieldlines (very few from localized region where horizontal fields undergone a permanent change, see Figure~\ref{Fig_Mos}(h-i)) become closed after reconnection at 23:48\,UT. During this reorganization, the stress in the coronal field releases and magnetic energy becomes redistributed from higher to lower heights, which contain a cusp-shaped current distribution and a closed field configuration.

\begin{figure*}[htb!]
\centering
\includegraphics[width=0.99\textwidth,clip=]{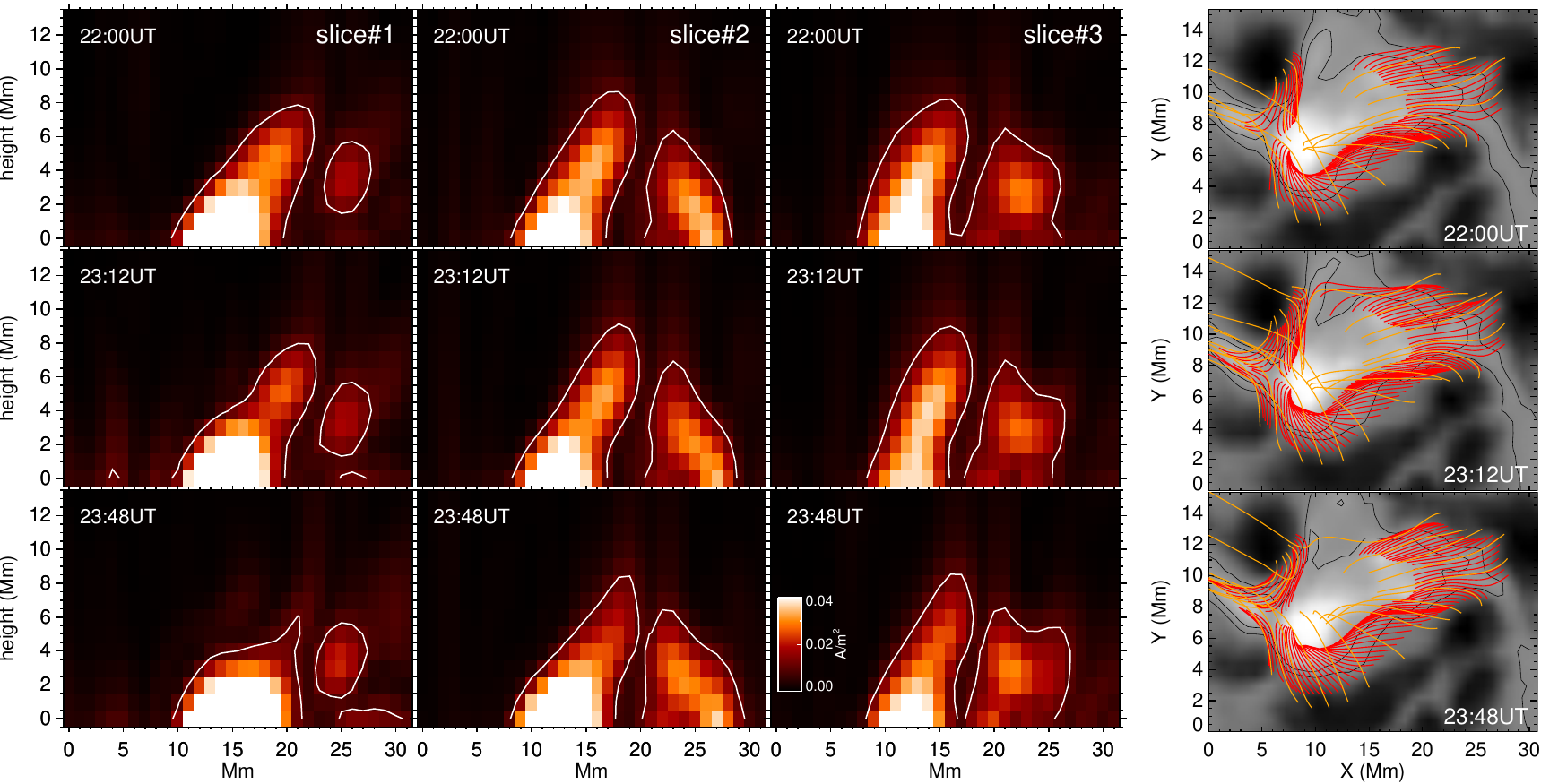}
\caption{Evolution of coronal current distribution in the planes of slices\#1, \#2, \#3 which are cross-sectioned (Figure~\ref{Fig_Isurf}) across the dome shaped fan structure before, during and after the flare. All panels are scaled within [0, 0.04]\,Am$^{-2}$ and contain a contour level of 0.01\,Am$^{-2}$. \textit{First column:} the slice plane is extracted across the PIL with unambiguous horizontal field change (Figure~\ref{Fig_Mos}(g-i)), pre-existing coronal currents build-up further to form stressed structures with a cusp-shaped current distribution (dissipating of these currents may drive reconnection), \textit{Second and third column:} Distribution of currents located somewhat outside the flaring region, which only slightly changed during the flare in comparison to their pre-flare distribution. Note that the two vertical concentrations of the currents in all panels correspond to a coronal separatrix layer intersecting with ribbon RB1. \textit{Last column:} The corresponding coronal field evolution above the sunspot SSP depicting the scenario for the coronal current generation. Fieldlines over the PIL are up to 1\,Mm height (red) and overlying lines (in orange) are up to 4\,Mm height. Note that some of the open fieldlines get closed at 23:48\,UT.}\label{Fig_Slj}
\end{figure*}

Following the procedure described in Section~\ref{Methods}, the distribution of Q on the photosphere (z=0) is calculated and compared with the photospheric vertical  current and the \Ha~ribbon locations (Figure~\ref{Fig_SqFac}). The computed values of Q are distributed in the range between 20 and $10^{13}$ with a grid size, eight times finer than the original model grid. At further fine resolutions the values of Q may still higher which is a computationally expensive task. It is inferred that when Q reaches the order of $10^6$, the thickness of the QSLs is small enough to allow reconnection \citep{demoulin1996, demoulin1997}. Therefore, the resolution at which we compute Q is small enough to locate important QSLs, which are taking part in reconnection. We show logarithmic Q maps, scaled within the minimum and maximum of the Q-distribution. The NLFFF model shows a more complex field structure compared with the PF, and consequently the Q pattern differs. The noisier transverse field, used as boundary condition for the NLFFF model contributes significantly to the patchy Q structure, contrary to noise-less smooth structures usually found in analytical configurations (e.g., \citealt{demoulin1996, titov2007}). To identify the important QSLs, which are relevant for the topology, \citet{savcheva2012} conjectured the significant QSLs to have both high Q-values and a high current density $|\textbf{J}|$. The authors used the product of Q and J to compare with flux rope structures. However, the JQ plots are not well proved to be identifying the main QSLs and the current density for potential field is zero; consequently we do not use them here. They also suggested to compute Q at different heights above photosphere to avoid patchiness.  
   
In both models of field extrapolations, the strongest Q values (black lanes of panels \ref{Fig_SqFac}(a-c)) form a structure like tad-pole in the strong field regions in addition to other patchy lanes related to noise, small magnitude, fragmented field across the AR. We notice the deviation of these small-scale pattern between PF and NLFFF models by the presence of sheared horizontal fields in the later. The high Q of this tadpole head is almost lying along the inversion line of magnetic field and also traces inner edge of the quasi-ellipsoidal ribbon RB1 (see panel \ref{Fig_SqFac}(g)). The tail of this tadpole Q pattern extends to RB2 through RB3. Note that the south-east part of RB3 does not have much relevance with RB1, however, the fluxes away from RB1 do have continuous mapping, as stated in earlier sections. Furthermore, the footprints of fan fieldlines (plotted in panels~\ref{Fig_Topo2}(a-b)) fall along this elliptical high Q which implies that the domed separatrix surface associated with the null is surrounded by a gradual connectivity gradient generalizing separatrices to QSLs. 

To compare these Q maps, we also plotted vertical current ($J_z$) in panel~\ref{Fig_SqFac}(i). The distribution of $J_z$ is in the range of $\pm150$\,mAm$^{-2}$, the higher values are located along PIL of the flaring region. Concurrently, the large values of the vertical current in the form of ridges around SSP also spatially coincide with this large elliptical Q values (especially head part of tad-pole). The AIA 94\,\AA~image, which captures the  hot coronal plasma loops ($\sim6$\,MK) is shown in panel~\ref{Fig_SqFac}(h). A comparison of Q map with the \Ha~flare ribbons is depicted in panel~\ref{Fig_SqFac}(g). 

Interestingly, there is a clear change in the trace of Q (marked by a blue arrow) in the head portion of the tadpole in the frames just before the flare at 23:00\,UT and after the flare at 23:48\,UT (panels~\ref{Fig_SqFac}(b-c)). The locus of Q transforms from a more zigzag path to a smoother one. Note that the change of this geometric path is more than one arc-second and it is in south part of the PIL, which is associated with the coronal field reconfiguration. The north part of the PIL also has a similar zigzag path, but does not change during the flare. Reorganization during magnetic reconnection releases high stress from the fields and therefore the path of Q and the QSLs change significantly around the flare time.

Moreover, the squashing degree distribution is evaluated in vertical cross section plane of tadpole head i.e., in the dome structure to construct coronal QSL shapes (see Figure~\ref{Fig_SqFac}(d-f). The overall shape of QSL follows fieldlines in dome structure with one of its legs located at one side of quasi-elliptical PIL and the second one is on other side, which are connected by a straight line across tadpole Q pattern. The middle section of this QSL is rooted close to the central part of SSP (as a downward spine) intersecting with the outer legs in the corona. Their shape also resembles with coronal current distribution (Figure~\ref{Fig_Slj}) obtained in similarly oriented planes. Specifically, the QSL extends to higher heights (maximum height is 6.8\,Mm) in NLFFF model than in the PF model (maximum height is 3.4\,Mm), because of currents in twisted fields. 

Interestingly, the maximum height of the QSL structure decreased by 2.37\,Mm during the evolution from 23:00\,UT to 23:48\,UT. This significant decrease in height is similar to the evolution of the coronal current distribution in slice\#1 of Figure~\ref{Fig_Slj}. The change in the height distribution of Q and $|\textbf{J}|$ is most likely caused by a reconfiguration of the field structure as explained earlier. With this characteristic finding, it is evidently clear that coronal QSLs are natural locations of currents, facilitating reconnection in them, whose signatures appear as ribbons in the lower chromosphere as a consequence of particle acceleration.

From this comparison, it is evidently clear that the domed-shaped fieldlines from ribbon RB1 are mapped to RB2 and RB3. Their association infers that the photosphere satisfies line-tied conditions, in which case the flare ribbons are created at the footpoints of QSLs and photospheric currents appear at similar locations of the flare ribbons as found here. Flare ribbons are generally locations defined by a high vertical current distribution \citep{vemareddy2012b} and their relation with QSLs (quantified by their Q-values) is topologically \citep{savcheva2012, janvier2014} found to be well associated. These observations imply a direct association of QSL, ribbons, and current distribution both at the photosphere and in the corona, as predicted from a standard flare models.

\begin{figure*}[!htp]
\centering
\includegraphics[width=.99\textwidth,clip=]{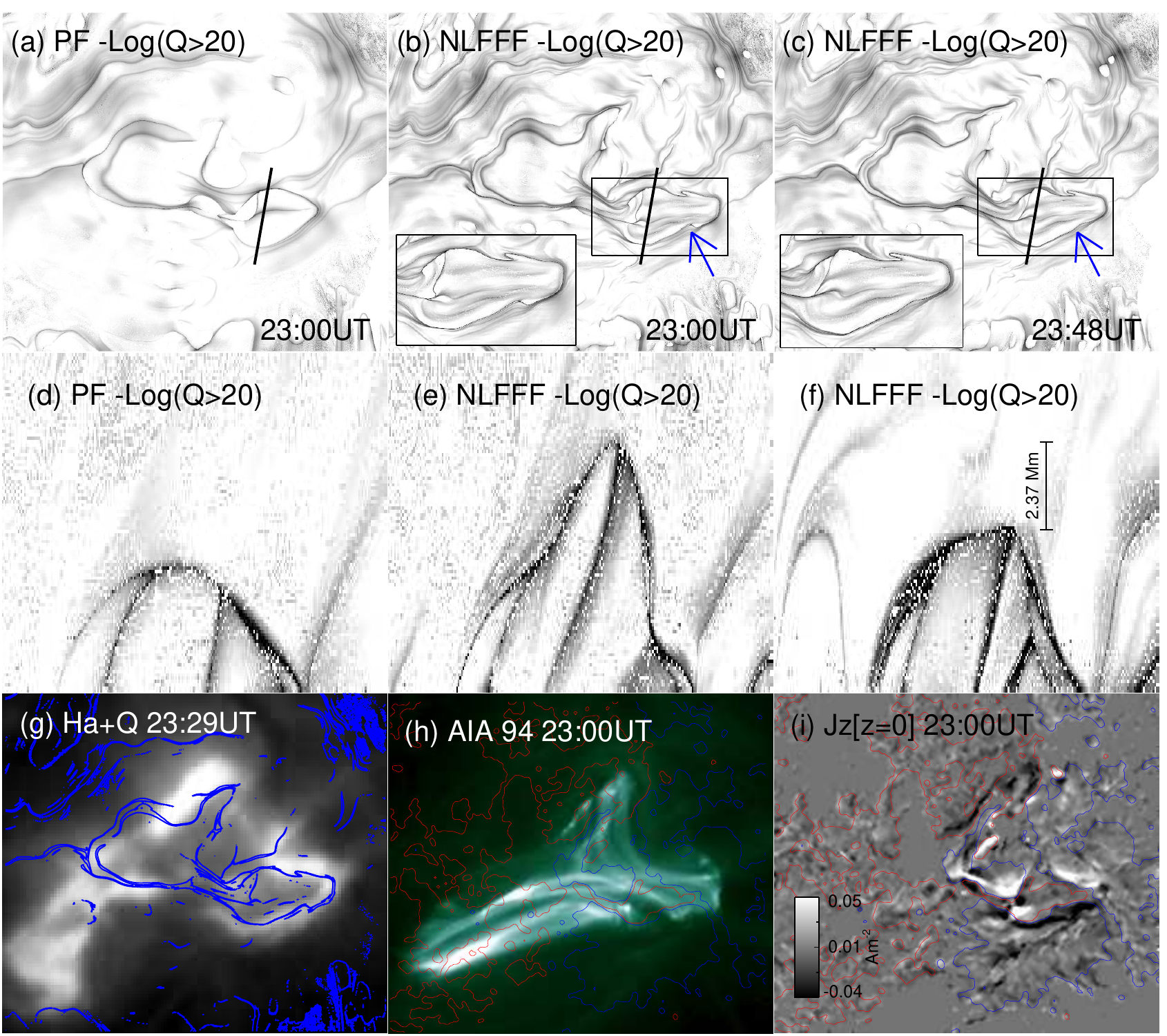}
\caption{Spatial distribution of squashing degree Q, which defines QSLs and its association with flare ribbons. Q at photospheric boundary ($z=0$) is calculated with (a) potential field (b-c) NLFFF approximations at, and displayed in logarithmic grey scale. Noticeably, high Q values form tad-pole like pattern in both models. The Q pattern differs at small scales especially at tadpole head portion in PF and NLFFF solutions.  An obvious change in Q between 23:00\,UT and 23:48\,UT  marked at the arrowed position is shown in enlarged insets, (d-f) The corresponding vertical cross-sections of Q in a plane are indicated by a solid line in panels (a-c). The maximum height of Q decreased by 2.37\,Mm from 23:00\,UT to 23:48\,UT. In all panels, the values of Q range from the lowest (in white) 20 to the highest $10^{13}$ but scaled appropriately for a better visibility, (g) An overlay of contours of Log(Q) on an \Ha~image illuminating the spatial association of Q and ribbons (h) AIA 94\,\AA~channel image with $B_z$ contours showing the coronal connectivity of ribbons and fluxes, (i) Spatial distribution of vertical current ($J_z$) also depict large values around positive flux sunspot SSP. Iso-contours of $B_z$ at $\pm150$\,G are drawn in red/blue. The FOV in panels (a-c), (g-i), is same as that indicated by white rectangular box (ROI) in Figure~\ref{Fig_Mos}.} \label{Fig_SqFac}
\end{figure*}

\section{Discussion}
\label{Disc}
Magnetic fields of the AR evolve quasi-statically under slow motions of line tied footpoints to the photosphere.  By modeling the AR magnetic field with a force free model, we studied the evolution of the magnetic structure covering the pre- and post phases of a flare. However, during the time of the flare, rapid changes of magnetic fields cannot be investigated with such an equilibrium model. Assuming a coronal Alfven velocity of 1000\,km\,s$^{-1}$, the time scale for Alfven wave traveling through the entire active region (300 arcsec) is less than 4 minutes. Consequently, dynamic features within this time scale cannot be reproduced in this model, as quasi-static condition for fields is not valid here. Nevertheless, by following the magnetic system evolution in the AR with a 12 minute time step, where quasi-static approximation remains valid, key information about the stability of the system e.g., free energy, and location of the current sheets can be obtained.}

Magnetic reconnection can be induced at null points with a fan-spine structure. Numerical simulations show that flux emergence and line-tied shearing boundary motions and rotational motions can induce current sheets near a 3D null point \citep{pontin2007,pariat2009, edmondson2010}. The flux systems below and above the fan are then able to interact near the null by magnetic reconnection.

The present confined flare event occurred in the magnetic configuration of an AR with a fan-spine topology, similar as studied earlier \citep{masson2009,xudong2012b,xudong2013}. The dome shaped fan separatrix is observed to intersect almost along quasi-elliptical ribbon RB1 and the spine fieldlines connect to remote ribbon RB3 (we assume this location belongs to RB3). Footpoints of fieldlines originating from the outer edge of the quasi-elliptical ribbon explain the extended regions of remote ribbons RB2 and RB3. Furthermore, within the six hour time span, there is no evident flux emergence from the flaring region. Nevertheless, flux motions upto 1.2\,km/s \citep{vemareddy2012a} are observed at this central part of the AR. These slow, continuous footpoint motions are the only supposed factors to build-up current layers in the coronal quasi/separatrices in the form of stress in fieldlines and they further induce reconnection at the coronal null. A predominant increase of the net current before the initiation of flare (see Figure~\ref{Fig_Wang}) in both polarity fluxes, but without appreciable flux emergence also supports this presumption. Also, time dependent MHD simulations for photospheric line-tying conditions \citep{karpen1990,karpen1991} confirmed the formation of quasi-static current sheets near separatrices. Therefore, these findings tend us to suggest that line-tied footpoint motions perturbed the fan-spine system inducing null point reconnection, resulting to flare emission at the footpoints of fan fieldlines and constitute quasi-elliptical ribbon.

The observations of this flare are very similar to that in \citet{masson2009}, except that flux emergence was identified in that study. Their interpretations of the prominent observational features with the aid of MHD simulations suit to our case and are useful to draw possible conclusions. Indeed, their simulation results helped \citet{wangh2012} to analyze five cases of circular flare ribbons, which are often associated with jets using solely \Ha~observations. Many observed features of flare ribbons may be difficult to explain with static models, due to the dynamic nature of the reconnection at the null point. Therefore, we will discuss them in connection with their simulation results.
 
During the reconnection, the fieldlines break and reconnect instantaneously at the separatrices or quasi-separatrices, slipping across each other. \citet{aulanier2006} demonstrated slipping (when fieldline velocities are sub-Alfvenic) and slip running (when fieldline velocities are super-Alfvenic) reconnection within the QSLs using numerical simulations. Evidence of this slipping reconnection was reported by \citet{aulanier2007} showing the coronal soft X-ray loops having opposite footpoint motions realizing slipping reconnection. 

When the reconnection phase commences at the coronal null point, emissions are expected to originate at the footpoints of underlying fan and spine fieldlines. According to the above simulation results, fieldlines closer to the null would reconnect first and that the slipping motion toward the null could then account for the counter-clockwise propagation of the circular ribbon emission. The same reasoning applies here for the emission of ribbon RB1. 

During sequential occurrence of slip and slip-running reconnection within the fan and null point reconnection, depending on the distance of the fieldlines from the spine, they slip or slip-run in directions and over distances that are compatible with dynamics of remote ribbon at the spine footpoints. As the observed remote ribbon RB3 is far away from main ribbon RB1 of null point reconnection, the delayed emission of RB3 by one minute can therefore be explained. The observed dynamics of ribbons are explained by slip and slip-running reconnection of fieldlines in QSL. However, there is no a priori reason to coexist both QSLs and separatrices with nulls in a given magnetic structure and connectivity. For that, the idea of separatrices embedded in larger QSLs was proposed by calculating gradients of connectivity with squashing degree \citep{masson2009}. We can see from Figure~\ref{Fig_SqFac} that the tadpole like pattern of very large values of Q ($10^{13}$, separatrices) is extended by intermediate values of Q (i.e., QSLs). Specifically in the region of tadpole head, the fan separatrix (large Q) facilitates null-point reconnection and the extended QSL (intermediate Q) regions allow fieldlines to slip and slip-running reconnection to explain the ribbon dynamics.   

After the impulsive phase of the flare, i.e., the plasma loops evolved to post flare arcades. This gives us the impression of disbursed fan-spine topology towards the relaxed state of potential fields. Because of this reason, the null point disappears in the frames after 23:36\,UT in either model (PF and NLFFF). When fieldlines are subjected to imposed boundary motions, which further brought them to a threshold level of sheared configuration, the fan-spine system collapses towards null point reconnection. Then these fieldlines are subjected to slip and slip-running reconnection. As a consequence, more fieldlines in the fan-separatrix along the southern part of the PIL become closed (Figure~\ref{Fig_Slj}). With this change in coronal field connectivity, the corresponding signatures such as decreased height of QSL in vertical cross-section, change in the footpoints of QSL at the boundary (Figure~\ref{Fig_SqFac}), and photospheric horizontal field enhancement (Figure~\ref{Fig_Mos}(i-h)) at a localized region are predominantly noted. 

\section{Summary}
\label{Summ}
We present a detailed study of an X1.5 flare using a force-free coronal magnetic field model. We focus especially on the coronal connectivity of the observed ribbon emission, caused by reconnection-accelerated particles toward lower layers of the solar atmosphere. We envisioned the existence of a particular type of topological structure associated with a magnetic null point. The existence of null points in the magnetic structure of the AR is of fundamental significance for the 3D-reconnection process. The main findings of this study are:

\begin{enumerate}
\item	The AR has an overall bipolar magnetic configuration with a small positive sunspot SSP surrounded by a group of small sunspots of negative polarity at the center. This central region forms an almost closed contour of the PIL and sheared horizontal fields are located along it.

\item	An X1.5 flare occurred, forming a quasi-elliptical \Ha~ribbon RB1 along and exterior of the closed PIL, and remote ribbons RB2, RB3. Emission in UV 1700\,\AA~spatially coincides with the \Ha~ribbons and have a peak emission at 23:23\,UT. Temporal and spatial association of HXR sources with these flare ribbons suggests that reconnection at the coronal null point above SSP is likely to have a fieldline linkage with ribbons, which is further ensured by coronal loop connections in hot 94\,\AA~observations. 

\item	Coronal PF and NLFFF models unveiled the presence of fan-spine topology associated with a single null point above SSP. The flare location of the HXR emission is well coincided with the null point location implying the  possibility of null point reconnection. Moreover, ribbon RB1 is almost co-spatial with the intersection of the fan-surface with the lower boundary, and outward spine connects to ribbon RB3. Fieldlines from outer edges of RB1 are linked to extended regions of remote ribbons RB2 and RB3, thereby explaining the topological relevance of observed ribbon emissions.  

\item  With the continued slow footpoint motions, shear stresses the fieldlines located along the PIL and leads to the formation of current layers (Figure~\ref{Fig_Isurf}) around the fan separatrix surface. These layers facilitate reconnection that dissipates coronal currents appearing as cusp shaped structures at lower heights. Reconnection releases energy and result in fieldlines close to a potential field.

\item At the lower boundary, the spatial distribution of fieldline connectivity gradients quantified as squashing degree Q have large values forming a tadpole like pattern. The head part of this pattern outlines the PIL and the inner edge of RB1, and the tail part extends to RB2 through RB3. Furthermore, the high Q values of this pattern is extending to intermediate values indicating separatrices of discontinuous connectivity embedded in QSLs of continuous fieldline connectivity gradients. These embedded separatrices in QSLs will facilitate slip and slip-running reconnection \citep{aulanier2006, masson2009}.

\item	While reconnection reorganises fieldline connectivity, the footpoints of separatrices/QSL at the boundary showed a transition from a zigzag path to a straight one in the south part of RB1. At the same time, the maximum height of the coronal QSL structure decreased by 2.37\,Mm during pre-to-post flare evolution. These significant changes suggest a clear topological change of the magnetic structure as the consequence of magnetic reconnection.

\item	With the knowledge from previous numerical studies \citep{masson2009, pariat2009} and the findings here, we suggest that line-tied footpoint motions perturbed the fan-spine system and caused null point reconnection, resulting to flare emission at the footpoints of the fan fieldlines lying in the quasi-elliptical ribbon RB1. During sequential reconnection at the null, the slipping motion of fieldlines toward the null explains the counter-clockwise emission of RB1. And slip and slip-running reconnection of fieldlines toward the outer spine can explain the one minute delayed emission of RB3 as compared to RB1.  
\end{enumerate}

Studying quasi-static evolution of coronal magnetic fields provides plenty of understanding on how the AR magnetic system stores and releases energy. It also provides details of the topological structure, which was appropriate for reconnection, and thereby explaining several aspects of the standard flare model. In other words, it successfully attempts to provide 3D details of 2D images of the flare seen in different wavelengths using a force-free coronal magnetic field model. This kind of modeling of the slow evolution of the AR magnetic fields is also useful to identify the gaps between observations and theoretical predictions, which can be tested by dynamical MHD simulations.    

\acknowledgements The data have been used here courtesy of NASA/SDO and HMI, AIA science team. We acknowledge the use of HPC cluster facility at Physical Research Laboratory. RHESSI is a NASA's small explorer mission. The author is grateful to Dr. Xudong Sun for useful discussions on the calculation of squashing factor. We thank an anonymous referee for his/her comments that improved the presentation of this manuscript. 

\bibliographystyle{apj}

\end{document}